\documentclass[pdflatex,sn-mathphys]{sn-jnl}
\usepackage{graphicx}

\usepackage[left]{lineno} 

\usepackage[utf8]{inputenc}
\usepackage{newunicodechar}
\newunicodechar{≥}{\geq}
\usepackage{tabularx}
\usepackage{booktabs}
\usepackage{enumitem}
\usepackage[colorinlistoftodos,textsize=small]{todonotes}
\usepackage{tabularx}
\usepackage{booktabs}
\usepackage{colortbl}
\usepackage{xcolor}
\usepackage{algorithm}

\usepackage[dvipsnames]{xcolor}

\usepackage{multirow}%
\usepackage{amsmath,amssymb,amsfonts}%
\usepackage{amsthm}%
\usepackage{mathrsfs}%
\usepackage[title]{appendix}%
\usepackage{xcolor}%
\usepackage{textcomp}%
\usepackage{manyfoot}%
\usepackage{algorithmicx}%
\usepackage{listings}%
\usepackage[numbers]{natbib} 

\theoremstyle{thmstyleone}%

\theoremstyle{thmstyletwo}%

\theoremstyle{thmstylethree}%

\begin{document}

\title[Article Title]{Predicting the risk of colorectal anastomotic leak based on preoperative mapping of the blood supply of the bowel}

\author*[1]{\fnm{Zahra} \sur{Tabatabaei}}\email{zata@di.ku.dk, elec.tabatabaei@gmail.com}

\author[1]{\fnm{Jon} \sur{Sporring}}\email{sporring@di.ku.dk}

\author[2,3]{\fnm{Mark} \sur{Bremholm Ellebæk}}\email{Mark.Ellebaek1@rsyd.dk}

\author[4]{\fnm{Alaa} \sur{El-Hussuna}}\email{alaanewemail@gmail.com}

\affil*[1]{\orgdiv{Computer Science Department}, \orgname{Københavns Universitet (KU)}, \orgaddress{ \city{Copenhagen},  \country{Denmark}}}

\affil[2]{\orgdiv{Department of Clinical Research}, \orgname{University of Southern Denmark}, \orgaddress{ \city{Odense}, \country{Denmark}}}

\affil[3]{\orgdiv{Research Unit of Surgery}, \orgname{Odense University Hospital}, \orgaddress{ \city{Odense}, \country{Denmark}}}

\affil[4]{\orgdiv{OpenSourceResearch Collaboration}, \orgaddress{\city{Aalborg}, \country{Denmark}}}

\abstract{Anastomotic leak remains one of the most serious complications following colorectal
cancer surgery, substantially affecting patient outcomes, recovery trajectories, and healthcare
costs. Despite advances in imaging technology, current preoperative assessment relies only on
clinical assessment, a process that is subjective, error-prone, and highly dependent on
individual expertise. To date, no validated CT-based method exists to predict anastomotic leak
risk prior to surgery. This protocol paper outlines a comprehensive framework for developing
and validating an AI-driven system for preoperative risk assessment using pre- and post-contrast CT imaging. The study describes the stages of data collection, ethical handling, and preprocessing of patient data in accordance with GDPR, image preprocessing, and the exploration of deep learning architectures designed to generate clinically interpretable outputs. Two integrated tools constitute the main deliverables of this workflow: 1) a risk assessment module, which quantifies the likelihood of leakage by analyzing vascular and tissue features in CT scans, and 2) a Content-Based Medical Image Retrieval (CBMIR) module, which identifies and displays similar historical cases to support evidence-based surgical decision making. The protocol paper requires close collaboration between hospitals and universities; this protocol demonstrates that such a system is technically feasible and clinically implementable within existing healthcare infrastructures. By following the proposed methodological stages and regulatory principles, other institutions can reproduce this workflow to develop analogous decision-support tools. Ultimately, this interdisciplinary framework aims to enhance surgical planning, reduce leak incidence, and contribute to a broader paradigm shift toward explainable,
data-driven precision surgery.}


\maketitle

\section{Introduction}\label{sec1}

Anastomotic Leak (AL) remains one of the most serious complications in colorectal
surgery, with rates of 2–7\% even in expert hands~\cite{slieker2012long,hyman2007anastomotic, Wu2025}. An anastomotic leak refers to the disruption or failure of a surgically created connection (anastomosis) between two segments of the gastrointestinal tract, commonly occurring after colorectal surgical operations~\cite{yung2024diagnostic, ganayee2025predicting}. A review of fourteen studies~\cite{khanPreoperativeAssessmentBlood2023} concluded that substantial variations in the arterial and venous supply of the colon and rectum may influence anastomotic leak rates. In addition, vascular calcification in the great blood vessels, which can be assessed using preoperative computed tomography, has been proposed as a potential predictor of anastomotic leakage~\cite{lie2024anastomotic}.

In current hospital practice, clinicians check the vascular tree before surgery to plan the anastomosis and optimize the chances for anastomosis healing [4]. This process is subjective, prone to human error, and depends on the level of expertise of the clinician. There are some diagnostic tools, such as CT scans, contrast enemas, and re-operation, that can help to confirm an anastomotic leak but cannot reliably predict its occurrence beforehand. This absence of reliable preoperative risk assessment limits clinicians’ ability to intervene preventively. Identifying high-risk patients for anastomotic leak enables preoperative optimization (e.g., correcting anemia, treating vascular disease, improving cardiovascular function), guides intraoperative choices such as stoma formation and anastomosis site, and ensures closer postoperative monitoring~\cite{grieco2020impact}.

Over recent decades, large volumes of medical imaging data have been collected. In
colorectal cancer care, preoperative CT scans are routinely acquired, creating a rich dataset that can support clinical decision-making if effectively utilized. Comparing a new patient’s scans with similar historical cases of anastomotic leakage (AL) can help clinicians make more informed assessments and anticipate complications. Advances in deep learning and shape analysis have enabled the development of retrieval tools that reduce human error and clinical workload. However, despite this progress, there is still no standardized or validated pipeline for CT-based preoperative risk prediction of colorectal AL\cite{taha2024machine}.

The proposed protocol seeks to bridge this gap by adapting the Content-Based Medical
Image Retrieval framework to surgical imaging and integrating it with a quantitative risk
assessment model, thereby moving toward a clinically deployable and explainable decision support system. In this paper, we present a protocol of a platform designed to substantially reduce the incidence of anastomotic leaks through improved risk stratification and evidence-based surgical planning. By enhancing preoperative assessment, the system aims to lower postoperative morbidity, mortality, and the need for permanent stoma formation. To achieve this, we propose a protocol of an AI-powered regression framework that analyzes paired pre-and post-contrast CT scans obtained before surgery to generate a calibrated, quantitative risk score. The vascular gastrointestinal tree serves as the central element of analysis, maintaining alignment with conventional diagnostic workflows while augmenting them through automation and quantitative interpretation. Furthermore, by incorporating information from surrounding mesenteric fatty tissues and perfusion patterns, the system captures additional imaging biomarkers that are often imperceptible to the human eye. These features are not only relevant for assessing anastomotic integrity but may also enable the evaluation of inflammatory changes in the mesentery, such as those observed in Crohn’s disease, and thereby broaden the clinical applicability of the method beyond colorectal cancer surgery.

In addition to the leak-risk estimator, we will explain the necessary steps to build a Content-Based Medical Image Retrieval (CBMIR) tool. This tool will present surgeons with
radiologically similar cases along with their corresponding surgical outcomes, enabling them to compare the query patient to prior cases. By reviewing clinical results associated with similar imaging patterns, clinicians can gain a broader perspective, assess risk more effectively, and analyze relevant similarities based on their medical expertise. Importantly, this approach adds an explainable, case-based layer of support rather than functioning as a black-box prediction system, thereby improving transparency and clinical trust. So, while surgeons today rely mainly on visual inspection of pre- and post-contrast CT scans, our method introduces an AI-based vascular analysis that provides an objective risk score. This innovation reduces subjectivity, supports surgical decision-making, and ultimately helps prevent anastomotic leaks. This project contains several high-risk elements that reflect its exploratory nature.

These tools will allow surgeons to better anticipate complications before surgery, reduce
the incidence of leaks, and share decisions with the patient. Figure 1 shows a graphical abstract of the proposed protocol paper. Accordingly, the proposed project highlights its exploratory nature and potential for high clinical impact. It is also worth noting that, while this work primarily focuses on anastomotic leak, it lays the groundwork for addressing other clinically relevant challenges, such as detecting small, visually imperceptible bleedings.

\begin{figure}[!htbp]
\begin{center}
\centerline{\includegraphics[width=0.90\textwidth]{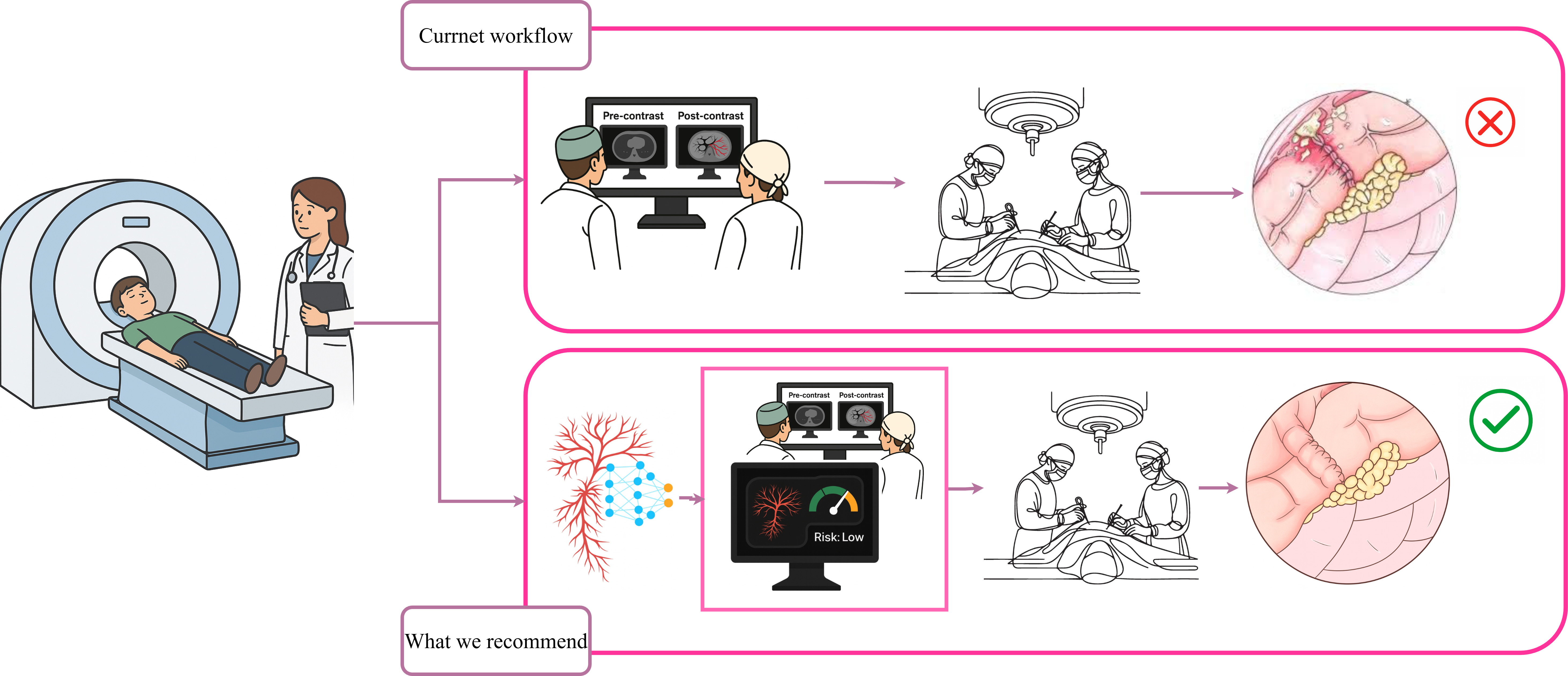}}
\caption{Comparison between the current clinical workflow and the proposed AI-assisted approach for surgical decision-making. The current method relies on visual inspection of pre- and post-contrast CT scans, often missing critical signs of complications such as anastomotic leakage. The recommended workflow incorporates a vascular-based AI risk estimator, providing surgeons with a quantitative leakage risk score, thereby enabling more informed preoperative planning and reducing the likelihood of post-surgical complications.}
\label{fig:abstract}
\end{center} 
\end{figure}

\section{Related work}\label{sec2}

Our protocol paper requires two complementary tools designed to assist clinicians: a CBMIR system and a risk assessment model. In this section, we review recent studies related to each tool to highlight their feasibility, existing challenges, and potential for clinical application.

\subsection{Content-Based Image Retrieval (CBIR)}

The CBIR framework has been widely adopted across various domains and imaging modalities. It has been applied not only to benchmark datasets such as CIFAR-10 and MNIST~\cite{ghalebContentbasedImageRetrieval2021, ImageRetrievalBased2022}, but also to more complex and clinically relevant modalities, including Magnetic Resonance Imaging (MRI) and CT~\cite{owaisEffectiveDiagnosisTreatment2019, kContentbasedMedicalRetrieval2025, ContentBasedMedicalImage2008, sudhishContentbasedImageRetrieval2024}. So, in the medical domain, it is named as Content-based Medical Image Retrieval (CBMIR). Furthermore, recent studies have extended the paradigm to histopathological imaging, demonstrating the capacity of CBMIR systems to handle high-resolution tissue samples and support diagnostic decision-making~\cite{tabatabaeiSiameseContentbasedSearch2025, wangRetCCLClusteringguidedContrastive2023, tabatabaeiAdvancingContentBasedHistopathological2024, kurmiContentbasedImageRetrieval2021}.

Across these studies, the primary objectives of CBMIR development include achieving high retrieval accuracy (typically measured as accuracy@K), extracting features that are both computationally efficient and semantically meaningful, and enabling multi-institutional connectivity through federated learning frameworks. For instance,~\cite{tabatabaeiWWFedCBMIRWorldwideFederated2023} introduced a distributed learning strategy that facilitates collaboration among medical centers without sharing raw patient data, ensuring compliance with privacy regulations such as GDPR.

Early CBMIR systems relied on handcrafted descriptors such as Scale Invariant Feature Transform (SIFT) and Local Binary Pattern (LBP) to capture local image texture and shape characteristics [18]. However, these approaches were limited in scalability and semantic representation. With the emergence of deep learning, convolutional autoencoders and Siamese networks have become dominant feature extractors, producing more discriminative and robust embeddings.

In the context of CT and X-ray imaging, transformer-based and hybrid models have recently demonstrated superior retrieval performance. For example,~\cite{saranyaMedicalApplicationDriven2025} proposed an advanced deep learning-based CBMIR pipeline to improve retrieval accuracy on complex X-ray image datasets. The method includes Hybrid Wavelet-Hadamard Transform (HWHT) for multi-scale detail enhancement, Gray-Level Co-occurrence Matrix for texture features, and a custom Transformer with global attention for similarity matching. On a COVID-19 chest X-ray dataset, their system achieved 99.5\% retrieval accuracy and 98.6\% on a pneumonia X-ray set. Kumar et al.~\cite{kumarVTHSCMIRVisionTransformer2024} proposed an end-to-end Vision Transformer Hashing framework with supervised contrastive learning, evaluated on chest X-rays, CT scans, and the BreaKHis histopathology dataset. Similarly, the hybrid CBMIR system proposed in~\cite{saranyaMedicalApplicationDriven2025} combined a hybrid wavelet–Hadamard transform, Bhattacharyya context global attention transformer, and sine chaos artificial rabbit optimization, achieving up to 99.5\% retrieval accuracy on COVID-19, pneumonia, and NIH CXR datasets. Another approach by Agrawal et al.~\cite{agrawalContentbasedMedicalImage2022} leveraged transfer learning-based deep models for disease-specific feature identification using standard COVID-19 chest X-ray datasets.

Souid et al.~\cite{souid2023improving} proposed a deep learning-based CBIR system to retrieve similar annotated images for lung disease diagnosis. Their method combines YOLOv5 for rapid lung finding localization with EfficientNet to extract robust features, allowing retrieval from a large chest radiograph and CT database. In the context of histopathological imaging, Núñez-Fernández et al.~\cite{nunez-fernandezContentBasedHistopathologicalImage2025} introduced a local–global feature fusion embedding model that combines local texture and global contextual features through multi-scale extraction, channel-attention fusion, and generalized mean pooling, achieving high accuracy on ImageNet-1k, PanNuke, and Kimia Patch24C datasets. Similarly, Tabatabaei et al.~\cite{MoreTransparentAccurate} presented a residual-block convolutional autoencoder that extracts compact latent embeddings, improving retrieval accuracy and speed. Other works, such as~\cite{mohammadalizadehNovelSiameseDeep2023}, explored hashing-based retrieval for breast cancer grading. In~\cite{golfeEnhancingImageRetrieval2025}, a Siamese network architecture inspired by the framework proposed in~\cite{tabatabaeiSiameseContentbasedSearch2025} was applied for multi-class grading of prostate cancer in the SICAPv2 dataset. These examples collectively demonstrate the versatility and diagnostic potential of CBMIR systems across diverse imaging domains.
Despite this progress, most CBMIR studies have focused on cancer detection, grading, and disease management across various modalities. To date, no study has applied CBMIR to surgical imaging or CT-based analysis for colorectal anastomotic leak risk prediction. This gap underscores the novelty of the proposed work, which adapts the CBMIR paradigm to pre- and post-operative CT data to support clinical decision-making in colorectal surgery.

\subsection{Risk assessment}
Prediction models are increasingly being used to support clinical reasoning and decision-making in modern medicine, particularly within the cardiovascular domain~\cite{RiskPredictionModels}. The study in~\cite{zhaoPredictingAnastomoticLeak2025} established and validated a nomogram to predict the occurrence of AL. A total of 231 eligible patients were divided into training and validation cohorts to feed into the univariate and multivariate analyses. The calibration curves resulting from these analyses showed good agreement between the predicted and actual AL occurrence.

Recent studies have moved to using deep learning and machine learning methods. The authors in~\cite{DevelopmentClinicalPrediction} investigated whether a predictive model that integrates multiple base models and patient-specific features can accurately estimate the risk of AL. The study included 9,120 patients and demonstrated that combining multisource information through a meta-model outperformed an individual base model. The meta-model achieved an F1 score of 87\% on cross-validation and 70\% on an external validation test set, highlighting the potential of advanced machine learning techniques to improve clinical decision-making in surgical care. In~\cite{DeltaCTRadiomicsBased}, the authors retrospectively analyzed the clinicopathological and radiological data of 213 patients with ESCC who received radical resection. 3D Slicer software was used in combination with Lasso extraction and 10-fold cross-validation to extract texture parameters from contrast-enhanced CT images and generate Delta-Rad scores. Several models were built using logistic regression to predict postoperative AL in ESCC. Therefore, the novel nomogram created using enhanced CT radiomics informed perioperative management and improved the survival quality of ESCC patients.

The study in~\cite{DiagnosticModalitiesEarly} advocates a multimodal approach for early leak detection by reviewing 33 papers. The authors in~\cite{DiagnosticModalitiesEarly}offered a tailored, patient-specific strategy to significantly enhance early leak prediction and management. It compiles evidence that postoperative serum C-reactive protein and procalcitonin are useful early biomarkers, while CT imaging can directly visualize leak signs such as fluid collections, pneumoperitoneum, or abscess in close proximity to the anastomosis. According to another review paper in this domain~\cite{MachineLearningDeep}, AI can provide surgeons with intraoperative feedback on blood supply and anatomical dissection planes, minimizing the risk of intraoperative complications and reducing the likelihood of AL development.

In addition to investigating risk predictor models, some studies focused on identifying the risk factors or parameters that affect AL or might cause AL. For instance,~\cite{taoAortaCalcificationIncreases2021} presented a study to evaluate whether aortic calcification is a risk factor for AL after gastrectomy in gastric cancer patients. They included 856 patients, among whom 818 did not have AL, and 38 cases had AL. In this study, they concluded that aortic calcification is an independent risk factor for anastomotic leakage after gastrectomy in gastric cancer patients. Similarly,~\cite{turhanImpactAorticCalcification2025} evaluated the association between preoperative aortic calcification and AL incidence while considering additional risk factors. According to~\cite{turhanImpactAorticCalcification2025}, aortic calcification as a marker of systemic atherosclerosis may impair tissue perfusion and anastomotic healing. Additionally, tumor factors (TNM stage, histology, and localization) and patient comorbidities (hypertension, cardiovascular disease, and neoadjuvant therapy) may contribute to AL risk. Severe aortic calcification is an independent predictor of AL in colorectal surgery. The authors believe that preoperative vascular assessments and comprehensive risk stratification models may help identify high-risk patients and guide perioperative management strategies to reduce AL incidence.

Although much work has been done on improving CTs, predicting AL\cite{pollmann2025preoperative}, and analyzing the important parameters of AL, there is still no study or tool using pre-/post-operative CT images that provides a quantified risk score. This is the gap that this study aims to fill through an interdisciplinary project, explaining all the required steps to demonstrate the feasibility of the approach by highlighting the key stages, regulations, limitations, and challenges. As illustrated in Figure 2, this work relies on a strong collaborative structure that integrates clinical expertise, advanced technical development, and real-world validation. Clinical partners provide access to high-quality CT datasets, surgical annotations, and expert knowledge necessary for defining relevant biomarkers and evaluating clinical utility. The technical research team focuses on the development of AI-based segmentation, image retrieval, and risk estimation models, ensuring methodological innovation and rigorous performance assessment. In parallel, international surgical experts contribute to broader clinical adoption by testing the system across different hospitals and patient populations, enabling the transition toward an open-source clinical application. Through this synergy between medical and computational domains, the project is positioned to deliver solutions that are both scientifically robust and practically deployable in diverse healthcare environments.

\begin{figure*}[!htbp]
\begin{center}
\centerline{\includegraphics[width=1\textwidth]{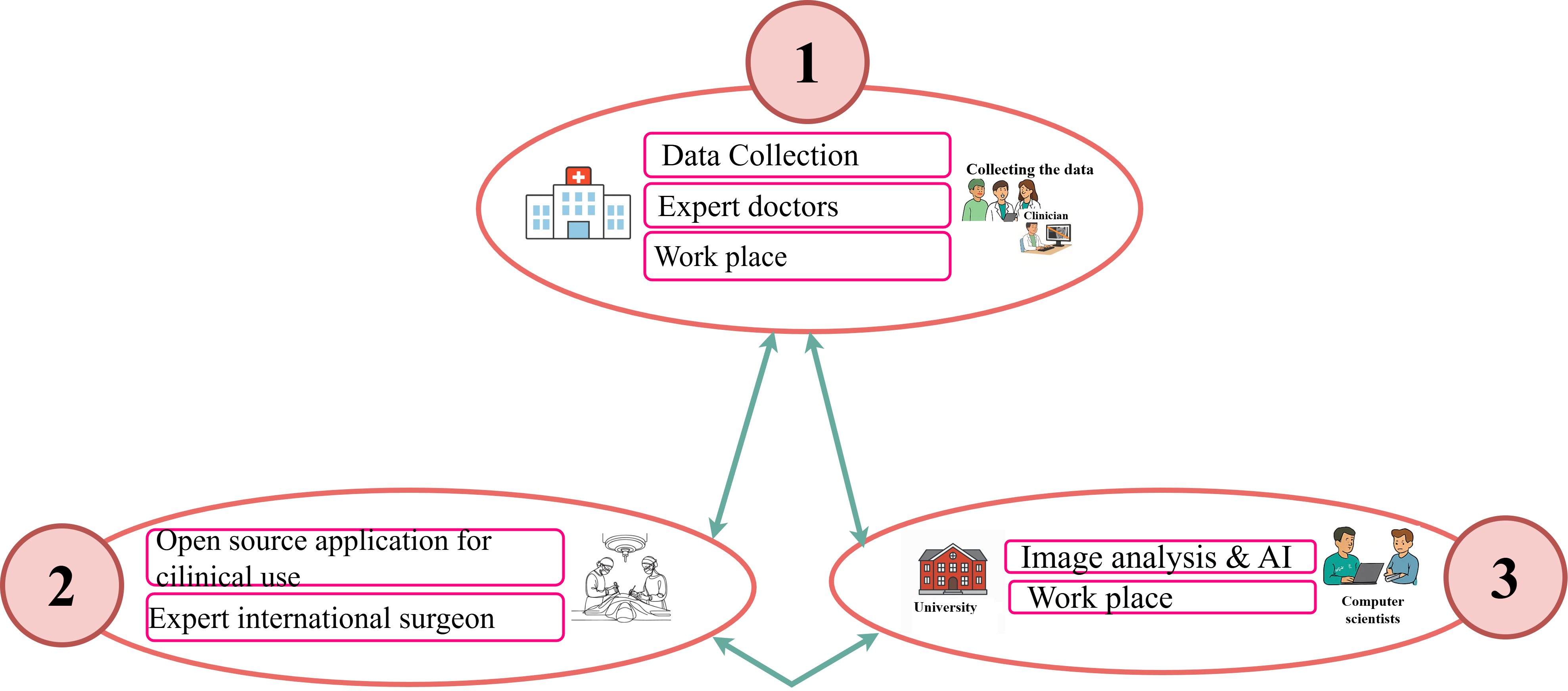}}
\caption{An interdisciplinary project pipeline.}
\label{fig:col}
\end{center} 
\end{figure*}

\section{Methodology}
This protocol requires several centers, including hospitals and universities, to collaborate as part of an interdisciplinary effort to collect medical data, preprocess it, and apply advanced deep/machine learning methods to extract meaningful features from CT images.

On the clinical side, the hospital offers unparalleled access to both expertise and data. Access to paired pre- and post-contrast CT scans, linked with surgical outcomes, ensures a clinically rich dataset that is aligned with ongoing activities at the hospital. This means that the infrastructure and regulatory environment at the hospital can provide the ideal setting for such a project. Secure data handling within hospital IT systems ensures GDPR compliance and patient confidentiality, while the integration of clinical researchers and data scientists in the same environment facilitates direct, iterative collaboration.

On the data science side, the central focus will be the development of a regression model for quantitative leak risk scoring and a retrieval tool for case-based decision support, and it requires the seamless integration of these complementary domains. This project leverages state-of-the-art deep/machine learning methodologies. The retrieval tool builds on topological data analysis and deep embedding methods to enable efficient and interpretable case comparisons. In parallel, an AI-powered regression model analyzes segmented vascular and colon features from preoperative CT scans to generate a calibrated risk score for anastomotic leak.

The team combines the expertise, infrastructure, and international outreach needed. Surgeons and radiologists define the clinical questions and validate outputs; data scientists and engineers design and refine the models, and the hospital offers both the data resources and the clinical pathways to translate research into practice. This close alignment across disciplines ensures that the developed regression platform and retrieval tool will not only advance scientific knowledge but also deliver tangible impact for patients and healthcare providers. By combining geometric descriptors, topological biomarkers, and deep network predictions, the model provides objective, quantitative estimates of complication risk. Uncertainty is assessed through data perturbation and ensemble techniques, ensuring stable and reliable predictions for clinical decision support. This methodological diversity reflects the interdisciplinary foundation of the team, which combines expertise in computer vision, medical AI, and mathematical modeling.

\begin{figure*}[!htbp]
\begin{center}
\centerline{\includegraphics[width=0.65\textwidth]{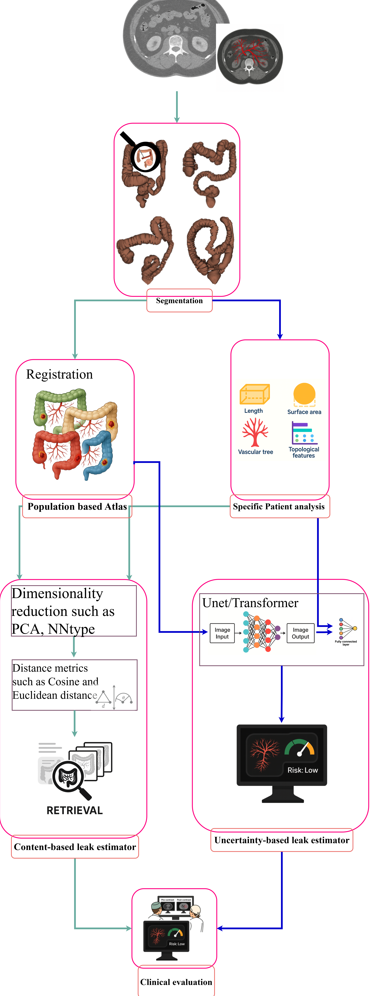}}
\caption{This protocol consists of the 1) curation of data, 2) segmentation of the anatomical structures, 3) registration of patients into a common atlas, 4) building of a library of shape features, 5) defining and comparing leak-risk estimators, and 6) development and testing of the system in a clinical setting.}
\label{fig:pipline}
\end{center} 
\end{figure*}

In the following sections, we describe how such interdisciplinary work can be carried out effectively and outline the A–Z primary responsibilities of each participating center, from collecting the data to working on the data and reaching the aim. Figure 3 illustrates each stage of this protocol.

\subsection{Data selection}
To collect the dataset, all patients who underwent colorectal surgery with a primary anastomosis over a defined period (e.g., 2010–2025) will be identified from a prospectively maintained database. All patients who received a postoperative CT scan due to a clinical suspicion of anastomotic leakage will be included, regardless of whether the suspicion was later confirmed or dismissed~\cite{marresImportanceRectalContrast2021}.
To ensure ethical and secure handling of clinical data, all CT volumes used in the study will be stored within the hospital. The computational workflow will be executed entirely within the hospital’s protected IT infrastructure, preventing any transfer of sensitive information outside the clinical environment. Before data access, personally identifiable information and sensitive metadata will be removed to guarantee full compliance with GDPR. Access privileges will be granted only to authorized personnel associated with the project, and system activity will be logged to maintain an auditable trail. Because the datasets consist of retrospectively collected and anonymized scans, the work is expected to fall under quality assurance regulations; however, the hospital will apply for ethical approval if required.
To determine an appropriate dataset size for model development, we performed a priori power analysis using both the ClinCalc Sample Size Calculator and GPower (version 3.1.9.4) to cross-validate assumptions and ensure robustness of the estimate. Based on published evidence indicating an anastomotic leak prevalence of approximately 5\% in colorectal surgery, we assumed a baseline incidence of 5\% and a clinically meaningful difference of 10\% in a high-risk subgroup. Using $\alpha$ = 0.05 and 80\% statistical power for a dichotomous endpoint, ClinCalc estimated a required sample size of 868 patients (434 per group), while GPower estimated 932 patients (466 per group) using Fisher’s exact test. As the two independent calculations yielded consistent results, it is recommended to include at least 1,000 pre-operative CT scans, ensuring a minimum of 50 confirmed leak cases for robust model development, training, and validation.

This dataset should be integrated with comprehensive clinical data, detailed records of surgical interventions, and postoperative outcomes, with a particular emphasis on the identification and characterization of AL. The time required to collect this dataset depends on hospital protocols; however, we estimate that approximately 5 to 6 months will be needed for data collection and data cleaning.

\subsection{Preprocessing Pipeline}

After secure access to anonymized CT data, a standardized preprocessing pipeline is applied to prepare the images for geometric and topological analysis. This includes normalization of voxel intensities to reduce variability introduced by scanner settings, acquisition protocols, and contrast injection timing. To isolate the cardiovascular structures of interest, segmentation algorithms are applied to separate vascular tissue from surrounding organs and background noise. The resulting vascular masks are cleaned to remove small spurious segments and are then converted into centerline representations that preserve geometric shape information while reducing data dimensionality.

To ensure consistency across scans, standardized resampling is applied to harmonize voxel spacing, and metadata checks are conducted to verify acquisition completeness. Quality control procedures are integrated to identify and exclude images affected by motion blur, insufficient contrast enhancement, or scanning artifacts. By filtering problematic scans early, we prevent error propagation and ensure that extracted geometric descriptors reflect reliable characteristics of vascular anatomy. The preprocessing pipeline also manages coordinate alignment, enabling consistent analysis across patients even when their imaging orientations differ.

Finally, the pipeline outputs a compact, structured representation of vessel geometry suitable for machine learning models. This representation includes centerlines, wall coordinates, and radius distributions that capture both local and global vessel morphology. By enforcing uniform preprocessing procedures, we ensure that downstream models operate on comparable inputs, which improves robustness, generalizability, and reproducibility of clinical risk prediction.

Depending on the number of images, we estimate that approximately 2 to 3 months will be needed to preprocess the collected data.

\subsection{Content-Based Medical Image Retrieval tool}

The CBMIR tool consists of two main components: offline and online stages. The offline stage includes a feature extractor that identifies and summarizes the most meaningful features from a collection of prior cases, thereby constructing a feature-based representation of the database. The input to this stage is a pool of previously acquired images, and the output is a compact database of their extracted features, which will subsequently be used during the online phase.

In the online stage, the same feature extractor is applied to a new query case. The extracted features of this query are then compared with those from the database using similarity measures such as Euclidean distance or cosine similarity. Based on these computations, the system ranks the prior cases according to their similarity to the query. Clinicians can then retrieve the top-K most similar images for comparison. The value of K depends on the clinician’s needs; it may range from a single similar case to the entire database. In most studies ~\cite{tabatabaeiAdvancingContentBasedHistopathological2024,tabatabaeiSiameseContentbasedSearch2025,wangRetCCLClusteringguidedContrastive2023, SMILY}, K is typically set to 3, 5, or 7; however, other implementations have used K = 10, 20, or even 400~\cite{mohammadalizadehNovelSiameseDeep2023}. 

According to the state-of-the-art studies ~\cite{mohammadalizadehNovelSiameseDeep2023, TopOCTopologicalDeep, TopologicalDataAnalysis, PersistentHomologyQuantitativea}, Siamese networks perform well on small datasets and achieve high accuracy in top-1 retrieval tasks ~\cite{tabatabaeiSiameseContentbasedSearch2025}. In addition, Topological Data Analysis (TDA) can extract topological features from images without requiring a training stage, thereby simplifying the feature extraction process for small datasets. Topological features of the images can be extracted by using a cubical complex, which extracts the number of loops and holes in each image and provides a topological representation for the image. The extracted features include geometric and topological descriptors such as centerlines, surface area maps relative to the anus, curvature and torsion of the colon, tree-based metrics of the vascular network, and spatial distances between the colon and vascular structures.

Building on this representation, a CBMIR platform will be developed to support efficient indexing and retrieval. Patient-specific features, along with their mappings to the atlas, will be compressed using various dimensionality reduction techniques, including PCA, t-SNE, and UMAP, as well as more advanced techniques such as minimal algorithmic information loss and correlation-preserving methods. The system will retrieve the most similar prior cases and define a leak-risk score based on the average observed leakage outcomes among the top-K retrieved cases.
The required time for designing a CBMIR tool depends on the number of images, the type of model, and the university’s infrastructure; however, we estimate that 7 to 8 months of investment time is required to design a CBMIR tool for this purpose.

\subsection{Risk estimator tool}

The goal of an accurate prediction model is to enable patient risk stratification, thereby supporting personalized clinical decision-making with the ultimate aim of improving patient outcomes and the quality of care~\cite{DevelopingPredictionModels}. To develop a reliable and precise risk estimation tool, the steps outlined in~\cite{PDFDevelopingClinical} can serve as a valuable guide.

The primary purpose of this tool is to provide clinicians with a risk score for anastomotic leakage based on CT imaging data. As such, its main users will be healthcare professionals. This tool can either be built upon existing models or developed from scratch, depending on the clinical variables selected for the prediction task.

This tool will take as input the segmented colon and vascular structures, registered to the shared atlas space. The aligned anatomical data, along with the original CT images and the geometric features extracted in the earlier stage, will be used to train both UNet and transformer-based models. These models will be designed to predict the likelihood of post-surgical complications, specifically focusing on anastomotic leakage.

To make the prediction clinically reliable, we will incorporate uncertainty quantification into the modeling framework. This will be done by introducing perturbations in input data and model weights, as well as by training ensembles of models initialized with different seeds. The predicted leak risk score will be defined as the average output across these perturbations or model ensemble members. This probabilistic risk score is intended to provide surgeons with calibrated and interpretable predictions. Finally, the entire system can be validated using newly collected cases in collaboration with surgeons.

The required time for designing a risk estimator tool depends on the number of images, the type of model, and the university’s infrastructure; however, we estimate that 5 to 6 months of investment time is required to design a CBMIR tool for this purpose.

\section{Perspective}

While the primary aim of this study is to develop a predictive framework for AL risk, the underlying methodology and tools offer significant promise for broader clinical applications. The integration of advanced AI models with geometric and topological features from CT imaging can be adapted to detect other subtle postoperative complications, such as small, otherwise imperceptible internal bleedings. This capability could significantly enhance early diagnosis and targeted interventions in critical care settings.

Moreover, the explainable CBMIR tool developed as part of this framework can support personalized treatment planning in oncology. For example, it can assist in the localization and comparative analysis of early-stage cancers by retrieving similar annotated cases based on shape, vascularity, and tissue characteristics. This opens new avenues for improving outcomes in local cancer detection and staging, particularly where human interpretation alone may be insufficient.

As part of future work, we plan to expand the applicability of the proposed approach to address other complex surgical challenges, such as identifying and localizing small internal bleedings that are difficult to detect through conventional imaging techniques.

\section{Conclusion}
The novelty of this project lies in the development of a leak risk estimator alongside a CBMIR tool. This will be accomplished by integrating advanced AI methodologies for segmentation, risk prediction, and explainability with innovative geometric and topological descriptors specifically designed for this medical domain. The effectiveness of the system will be validated in real-world clinical settings through close collaboration with surgeons.

\section{Funding }
This research did not receive funding.

\section{Competing interest}
We declare that the authors have no competing interests as defined by BMC or other interests that might be perceived or influence the results or discussion reported in the article. 
\section{Author contribution}
Zahra Tabatabaei: Writing—original draft, review \& editing, visualization, investigation, methodology, formal analysis, data curation, conceptualization.

Jon Sporring, Mark Bremholm Ellebæk, and Alaa El-Hussuna: review \& editing, investigation, and conceptualization.

\section{Data availability, Ethics}
Not applicable

\bibliographystyle{elsarticle-num}   
\bibliography{refs}    
\end{document}